\documentclass{article}
\usepackage{subfigure,url}
\usepackage{amsmath}
\usepackage{amssymb}
\usepackage[ruled,vlined]{algorithm2e}
\usepackage{hyperref}
\usepackage{setspace}
\usepackage{color}
\usepackage{graphicx}
\usepackage{authblk}

\newcommand{\card}[1]{\ensuremath{|#1|}}

\newcommand{\degree}[1]{\ensuremath{\delta_{#1}}}
\newcommand{\maxdegree}{\ensuremath{\Delta}}

\newcommand{\coloring}{\ensuremath{C}}
\newcommand{\Nset}{\ensuremath{\mathbb{N}}}

\begin{document}

\title{On Distributed Graph Coloring with Iterative Recoloring}

\author[1,2]{Ahmet Erdem Sar{\i}y\"uce\footnote{Corresponding author: sariyuce.1@osu.edu. Tel: 614-688-9637, Fax: 614-688-6600}}
\author[4]{Erik Saule}
\author[1,3]{\"Umit V. \c{C}ataly\"urek}
\affil[1]{Department of Biomedical Informatics, The Ohio State University}
\affil[2]{Department of Computer Science and Engineering, The Ohio State University}
\affil[3]{Department of Electrical and Computer Engineering, The Ohio State University}
\affil[4]{Department of Computer Science, University of North Carolina at Charlotte}

\maketitle

\begin{abstract}
  Identifying the sets of operations that can be executed simultaneously
  is an important problem appearing in many parallel applications.  By
  modeling the operations and their interactions as a graph, one can
  identify the independent operations by solving a graph coloring
  problem. Many efficient sequential algorithms are known for this
  NP-Complete problem, but they are typically unsuitable when the
  operations and their interactions are distributed in the memory of
  large parallel computers. On top of an existing
  distributed-memory graph coloring algorithm, we investigate two
  compatible techniques in this paper for fast and scalable distributed-memory
  graph coloring. First, we introduce an improvement for the
  distributed post-processing operation, called recoloring, which
  drastically improves the number of colors. We propose a novel and efficient
  communication scheme for recoloring which enables it to scale
  gracefully. Recoloring must be seeded with an existing coloring of
  the graph. Our second
  contribution is to introduce a randomized color selection strategy for initial
  coloring which quickly produces solutions of modest quality. We extensively evaluate the
  impact of our new techniques on existing distributed algorithms and show
  the time-quality tradeoffs. We show that combining an initial
  randomized coloring with multiple recoloring iterations yields
  better quality solutions with the smaller runtime at large scale.
\end{abstract}

\section{Introduction}

In parallel computing, the problem of organizing
computations so that no two concurrent procedures access shared
resources simultaneously appears often. This problem might be solved by ordering
them explicitly so that concurrent accesses can not happen, using
concurrency control mechanisms such as locks, lock-free data structures
or transactional memory. Even when using concurrency controls
mechanisms, it is important to minimize the access to the lock or page
of the transactional memory. The problem can be modeled as a graph
coloring problem where the vertices of the graphs are operations of
the problem and edges represent concurrent accesses to a resource. A
coloring of the graph is a partition of the vertices of the graph in a number of independent
sets. Minimizing the number of colors (i.e., the number of independent sets) reduces the number of
synchronization points in the computation and enhances the efficiency
of the parallel computers. 

Graph coloring appears in many other applications, including, but not
limited to, optimization~\cite{CM83}, efficient computation of sparse Jacobian and Hessian
matrices~\cite{GMP05}, preconditioners~\cite{Saad99}, iterative solution of sparse linear
systems~\cite{Jones94}, sparse tiling~\cite{Strout02}, printed circuit
testing~\cite{Garey_circuit}, eigenvalue computation~\cite{Manne98aparallel}, frequency assignment~\cite{gamst_freq}, parallel numerical computation~\cite{ABC94} and register allocation~\cite{Chaitin} areas.

There exist different types of the graph coloring problem. For instance, 
the distance-$k$ coloring problem requires that vertices separated by less 
than or equal to $k$ edges to have different colors.
In this paper, we are only interested in distance-1
coloring; however, we believe that all the techniques and results presented
in this paper can be extended to the other variants of the graph
coloring problem.

Let $G=(V, E)$ be a graph with $\card{V}$ vertices and $\card{E}$
edges. Set of neighbors of a vertex $v$ is $adj(v)$; its
cardinality, also called the {\em degree} of $v$, is \degree{v}. The
degree of the vertex having the most neighbors is $\maxdegree = \max_v
\degree{v}$. A distance-1 coloring $\coloring:V \rightarrow \Nset$ is a function
that maps each vertex of the graph to a color (represented by an
integer), such that two adjacent vertices get different colors,
i.e., $\forall (u, v)\in E, \coloring(u) \neq \coloring(v)$. Without
loss of generality, the number of colors used is $\max_{u\in V}
\coloring(u)$. Finding a coloring with as few colors as possible
is an optimization problem. The problem of deciding whether a graph
can be colored with less than $k$ colors is known to be NP-Complete for
arbitrary graphs~\cite{matula_SL}. Therefore, finding the minimal
number of colors a graph can be colored with (also called the
chromatic number of the graph) is
NP-Hard.  Recently, it has been shown that, for all $\epsilon > 0$, it
is NP-Hard to {\em approximate} the graph coloring problem within
$\card{V}^{1-\epsilon}$~\cite{Zuckerman}. Hence, we are focusing on 
heuristics which work well in practice on many graphs.

We are studying the distributed-memory graph coloring problem in this paper.
We focus on obtaining high quality solutions for large-scale scientific parallel
applications. In those applications, the computational model (the graph)
is already distributed to the nodes of the parallel machine. 
If the graph is sufficiently small, with a naive approach, one can aggregate it in the
memory of a single node and color it there. It would actually be better to take
advantage of partitioning in this case, i.e., first coloring the {\em interior}
vertices (vertices for which all their neighbors are local) in parallel and
then color the other, {\em boundary vertices}
(vertices that have at least one non-local neighbor), sequentially, by aggregating
the graph induced by them in the memory of a single processor. 
However, even if one implements such optimizations, one can still
achieve significantly faster solutions by coloring in parallel~\cite{BGMBC-jpdc}.
If the graph is too large to fit in the memory of a single computer,
coloring in distributed memory is inevitable. Also, people
usually use coloring as a tool in contexts where good and quick solutions are desired.
Therefore, one cannot afford repartitioning the graph for the sole purpose of coloring, since
repartitioning has a higher computational complexity than coloring
itself.

In this work, we aim to achieve good quality coloring with good
runtime in distributed-memory settings. As a starting point, we
use the framework presented in~\cite{BGMBC-jpdc}; it mainly relies on
processor-local greedy coloring techniques with multiple iterations to
converge to a valid solution. Traditionally, the problem of improving
the quality of coloring obtained from greedy coloring methods is
addressed by considering the vertices in an order that have good properties for
coloring.  We investigated such techniques in~\cite{HiPC11} and found
that processor-local ordering techniques do not yield quality
improvement at large scale. Good global ordering techniques
exhibit little parallelism and will not yield good runtimes.  In this
paper, we investigate recoloring, a post-processing operation which
refines an existing coloring. We showed in~\cite{HiPC11} that the quality of the
solution the recoloring procedure generates is not affected by the
scale of the distributed-memory machine. However, the communications
required by the algorithm make it non-scalable in terms of the runtime.

In this paper, we show how recoloring can be made scalable in terms of
runtime by improving its communication scheme. In particular, we use
piggybacking to reduce the number of communications. With a fast
recoloring method, it becomes efficient to run recoloring multiple
times. This allows to potentially start with a faster initial coloring algorithm of
lower quality. Pursuing this idea, we investigate the Random-X Fit
color selection strategy~\cite{Gebremedhin02paralleldistance-k} for
generating a first coloring. Random-X Fit leads to a solution of
modest quality but with a balanced color distribution which makes it very
suitable for recoloring. We show that combining these different
techniques allows to reach better time-quality trade-offs than
previously existing algorithms. We extensively evaluate all the
techniques we propose at different scales in terms of both quality and
runtime. This allow us to identify two sets of parameters ``speed'' and ``quality''
which enables the user to achieve the tradeoff she is interested in without having
to understand the inner working of our coloring framework.

The remaining of the document is organized as
follows. Section~\ref{sec:rel_work} presents the different coloring
algorithms existing for sequential and parallel
architectures, including the previous 
distributed-memory coloring algorithm which is the reference algorithm we
use and ordering solutions. Section~\ref{sec:improv} discusses the techniques we use to improve
coloring in distributed-memory architecture. The proposed techniques
are experimentally evaluated on real-world graphs and on random graphs
in Section~\ref{sec:expe}. Final conclusions and ideas to improve further
are given in Section~\ref{sec:ccl}.

\section{Graph Coloring Algorithms}
\label{sec:rel_work}

Graph coloring is one of the well studied problems in the
literature~\cite{Ellis:1989:LVG:74142.74153,dubr81,GMP05,kosow}.
Literature is abundant with many different techniques, such as the one
that utilizes greedy coloring~\cite{matula_72, kosow},
cliques~\cite{Turner:1988:AKC:48880.48884} and Zykov
trees~\cite{dubr81}. In the following sections, we will first present
a simple greedy sequential coloring algorithm and how it can be
improved with vertex visit orderings and recoloring. We then briefly
discuss other shared- and distributed-memory parallel coloring
algorithms.

\subsection{Sequential Coloring}

In spite of the existing pessimistic theoretical results and the existence of more
complicated algorithms, for many graphs that shows up in practice,
solutions that are provably optimal or near optimal can be obtained
using a simple {\em greedy} algorithm~\cite{CM83}. In this algorithm,
the vertices are visited in some {\em order} and the smallest
permissible color at each iteration is assigned to the vertex. 
Algorithm~\ref{a:greedy} gives the pseudocode of this technique.

\begin{algorithm}
\DontPrintSemicolon
\SetKwComment{tcp}{$\triangleright$}{}

\caption{Sequential greedy coloring.}
\label{a:greedy}
\KwData{${G = (V, E)}$}

  \For{{\bf each} $v \in V$} {
      \For{{\bf each} $w \in \mathit{adj}(v)$} {
      \textsf{forbiddenColors}[\textsf{color}[$w$]] $\leftarrow$ $v$ \;
}
      \textsf{color}$[v]\leftarrow \min\{i>0: \textsf{forbiddenColors}[i] \neq v\}$ \label{l:assign} \;
}
\end{algorithm}

Different ways for improving the resulting number of colors
are presented in the literature. Changing the color selection strategy
has an impact in the quality of coloring. Choosing the smallest permissible color,
as stated in Algorithm~\ref{a:greedy},
is known as the {\em First Fit} strategy. Selecting a color based
on an initial estimate of the number of colors in a distributed-memory setting
is proposed in~\cite{BGMBC-jpdc} and called {\em Staggered First Fit}.
The {\em Least Used} strategy picks (locally) the least used color so far so that
a more even color distribution is achieved. Apart from that, 
Gebremedhin et al.~\cite{Gebremedhin02paralleldistance-k}
proposed {\em randomized} color selection strategies.

Algorithm~\ref{a:greedy} has two nice properties. First, for any
vertex-visit ordering, it produces a coloring with at most
$1+\maxdegree$ colors. Second, for some vertex-visit orderings it will
produce an optimal coloring~\cite{GMP05}. Many heuristics for ordering
the vertices have been proposed in the literature~\cite{GMP05}. These
heuristics can be grouped into two categories: {\em static} and {\em
  dynamic} orderings.  {\em Largest First} (LF) and {\em Smallest
  Last} (SL)~\cite{Matula1983} orderings are {\em static} orderings,
in the sense that the coloring order is obtained before the coloring
starts. {\em Saturation Degree}~\cite{Brelaz1979} and {\em Incidence
  Degree} orderings are {\em dynamic} orderings in which the coloring
order of the vertices is obtained while the coloring is done. We refer
the reader to~\cite{GMP05} for a summary of these ordering
techniques. The LF ordering, introduced by Welsh and
Powell~\cite{Welsh01011967}, visits the vertices in non-increasing
order of their degree. SL obtains the ordering from backwards via selecting a
vertex with the minimum degree to be ordered last and removing it from
the graph for the rest of the ordering phase. Then, the next vertex
with the minimum degree within the remaining graph is selected to be
ordered second-to-last. This procedure is repeated until all vertices
are ordered. 

Apart from the different pre-ordering techniques,
Culberson~\cite{Culberson92iteratedgreedy} introduces a coloring
algorithm, called {\em Iterated Greedy} (IG), where starting with an initial
greedy coloring, vertices are iteratively \textbf{recolored} based on the
coloring of the previous iteration. Culberson shows that, if the
vertices belonging to the same color class (i.e.,
the vertices of the same color) at the end of initial coloring are colored
consecutively, then the number of colors will either decrease or stay the
same. Several different permutations of color classes are considered
based on the colors of the vertices in the previous iteration, the
number of vertices, degree of the vertices, randomness
and combination of them. Culberson suggests that a hybrid approach, which
changes the permutation strategy at each recoloring iteration is more
effective than applying the same permutation strategy at each round. He proposed using 
random permutations of the colors to break cycles where number of colors stays the same after
some number of iterations that appear when deterministic methods are used exclusively.
The first work evaluating the effects of the recoloring scheme on parallel 
graph coloring is~\cite{Gebremedhin_parallelgraph} where the focus is
on shared-memory computers. Also, Goldberg et al. suggest to use recoloring
in multigraph edge-coloring~\cite{JGT:JGT3190080115}.

\subsection{Distributed-Memory Parallel Coloring}
\label{sec:jpdc}

In the literature, Bozda\u{g} et al.~\cite{BGMBC-jpdc} introduced the first
scalable distributed-memory parallel graph coloring framework. We believe that, their
work is the only distributed-memory coloring algorithm that provides
a parallel speedup, therefore our work is based on their algorithm, which we
present here briefly.

It is assumed that the graph is distributed onto the distributed
memory of computers. Each vertex belongs to a single processing unit. 
At each processing unit, information of the edges connected to any owned vertex
is kept. In other words, for an edge $(u, v)$, if both $u$ and $v$ belong to
a single processing unit, then only this processor has the information of that edge.
Given the edge $(u, v)$, if $u$ and $v$ belong to two different
processing units, then $(u, v)$ is only know by these two processing units.
Each processing unit colors the vertices it has. If a vertex $u$ has a neighbor
vertex $v$, owned by another processing unit, then $u$ and $v$ are {\em
  boundary} vertices. On the other hand, if all neighbors of a vertex $u$
belong to same processing unit with $u$, then $u$ is called as {\em internal} vertex.

The coloring procedure is performed in multiple rounds. All the colorless vertices are
tentatively colored by the greedy coloring algorithm at each round.
Two neighbor vertices, belonging to two different processing units, may
end up having the same color at the end of this phase, i.e., a conflict can occur.
Such conflicts are independently detected by each processing unit. To resolve a conflict,
one of the vertices will keep its color whereas the other one is scheduled to be
colored again in the next round. Ties are broken based on a random total ordering,
which is obtained beforehand. The algorithm continues to iterate in multiple rounds until there is no conflict.

Each round of the coloring procedure is executed in supersteps to reduce the number of conflicts.
Each processing unit colors a specified number of owned vertices at
each superstep. After that, colors of the boundary vertices (colored in that super step)
are exchanged among {\em neighbor} processing units. There are two types of
coloring procedures based on the communication mechanism between processing units.
Coloring is called as {\em synchronous}, if a processing unit waits for its
neighbors to complete their super step before beginning the next one. {\em Synchronous}
coloring guarantees that two vertices can be in conflict only if they are colored
in the same superstep. The other option, in which no waiting mechanism is enforced, is named
as {\em asynchronous}. The superstep size matters for the coloring procedure, since a larger
size decreases the number of exchanged messages with likely high number of conflicts, whereas
the smaller size increases the exchanged message number with expectedly low number of conflicts.

\subsubsection{Vertex-visit Ordering}

The orderings considered in~\cite{BGMBC-jpdc} only consider the 
partitioning of the graph and not the properties of the graph
itself. Three orderings were investigated in their work, coloring
internal vertices first, boundary vertices first and coloring the
vertices in the order they are stored in the memory (which was called
{\em unordered} in~\cite{BGMBC-jpdc}, here we will call it {\em Natural} ordering).

In~\cite{HiPC11}, we investigated two more successful ordering heuristics,
namely {\em Largest First} (LF) and {\em Smallest Last} (SL)
techniques.  The LF ordering can be computed in $O(\card{V})$ time~\cite{Welsh01011967}.
Using a carefully designed bucket data structure allows to implement SL with a
complexity of $O(\card{E})$~\cite{Matula1983}. In distributed memory, we let each
processor compute an ordering of the graph based on the knowledge it has and
therefore it is not guaranteed that the coloring computed on a single processor and on multiple
processors will be the same. We also verified this fact by scalability experiments
in~\cite{HiPC11}. We showed that LF ordering provides less number of colors than
sequential Natural ordering on less than 8 processors. SL obtains much fewer number
of colors. However, when the number of processors increases, the advantages of
LF and SL orderings disappear. On 512 processors, the choice of vertex-visit
ordering does not yield a significant difference in number of colors. So, vertex-visit orderings
are not beneficial for quality at large scale since the number of internal
vertices decreases when the scale increases and the ordering is done locally at each processor.

\section{Distributed-Memory Iterative Recoloring}
\label{sec:improv}

Culberson~\cite{Culberson92iteratedgreedy} presented the use of
iterative recoloring for improving the number of colors in a
sequential algorithm. In~\cite{HiPC11}, we extended this work to distributed-memory
graph coloring. The recoloring idea naturally fits distributed-memory
graph coloring.  Independents sets obtained at the end of the initial
coloring are used to efficiently color the graph. Our {\em recoloring algorithm} (RC)
proceeds in as many steps as the number of colors in the initial
solution. All the vertices in the same color class (i.e., having the
same color in the previous coloring round) are colored at the same
step in the recoloring process. Then the processors exchange the color
information with their neighboring processors at the end of the
step. Notice that, although a processor does not have any vertices with
that color, it will wait for other processors to finish coloring at that
step. This procedure guarantees that no conflict is created by the end of
recoloring. Therefore, recoloring in
sequential and in distributed memory lead to the same solution, if
the initial coloring is same, making
the recoloring scalable in terms of number of colors.

However, the algorithm is fairly synchronous since a processor can not
start the $i$-th step before its neighbors finish their $(i-1)$-th
step. Moreover, there is no guarantee that two processors will have a
similar number of vertices in each color, thus potentially leading to
a load imbalance. To prevent some potential load imbalance due to
synchronous execution of RC, we also propose a second recoloring
approach, named {\em asynchronous recoloring} (aRC). In this
approach, each processor computes their vertex-visit
orderings independently using the initial coloring, and apply a second parallel
coloring with this new vertex-visit ordering using the algorithm
recalled in Section~\ref{sec:jpdc}. Note that, conflicts can
happen at the end of this procedure, so this second parallel
coloring step proceeds in conflict resolution rounds. We expect
that this approach will not be as good as synchronous recoloring in
terms of number of colors, but it might be beneficial for trading the
quality for runtime.

In the recoloring process, all the vertices in the same
color class must be colored in a consecutive manner, but one can
choose any permutation of the color classes and this permutation affects
the number of colors well. Therefore different permutations
of the color classes should be taken into account for sequential and parallel
recoloring. We considered four permutations of color classes: {\em
  Reverse} order (RV) of colors, {\em Non-Increasing} number of
vertices (NI), where the color classes are ordered in the
non-increasing order of their vertex counts, {\em Non-Decreasing
  number} of vertices (ND), which is similarly derived, and {\em
  Random} order of color classes where the color classes are ordered
randomly using the Knuth shuffling procedure in linear
time~\cite{Culberson92iteratedgreedy}.  We compute global NI and
ND permutations by communicating between processors before coloring to
exchange the number of vertices at each color class.

\subsection{Improvements on the Communication Scheme of Synchronous Recoloring}
\label{sec:improved_comm}

The communication scheme in {\em synchronous recoloring} algorithm
results in fine-grained communications where the number of messages
are high but each of these messages is small. This behavior lowers
network performance and results in slower execution. To improve this
fine-grained communication scheme, we apply piggybacking techniques.

In {\em synchronous recoloring}, vertices belonging to same color
class are colored in the same step and colors are communicated at the
end of that step immediately. However, immediate communication is not
necessary for two boundary vertices that are not colored in
consecutive steps. For example, for two neighbor boundary vertices $a$
and $b$, belonging to different processors, if $color\_order(a)
\textgreater color\_order(b)$ (in other words if $a$ will be colored
later than $b$), then the coloring of $b$ does not depend on the
coloring of $a$. Of course $b$ will need the new color of $a$ during
the next recoloring iteration, so sending the color of $a$ to $b$'s
processor can be deferred to the end of the recoloring iteration.  For
the reverse case, a similar technique can be used. Assume again that
$color\_order(a) \textgreater color\_order(b)$, then $a$ needs to know
the new color of $b$ before being colored. So, sending the color of
$b$ can happen at any time between step $color\_order(a)$ and
step $color\_order(b)$.  A processor $P_1$ accumulates the color
information in a buffer to send to a processor $P_2$. $P_1$ only
sends the whole buffer at the color step before the step where $P_2$
needs any of the information contained in the whole buffer. This way, the color
information of vertices are piggybacked and sent in a minimum number
of messages which improves the performance of the communication subsystem.

In order to be able to apply the piggybacking technique, each
processor needs to have the knowledge of when and from whom
to receive messages for each iteration. For this purpose, there is a need for
pre-communication at the beginning of each recoloring iteration among
processors so that each get the information of from whom
to receive at which step. The only information processors should
send to each other is the list of color classes when they will
communicate with each other. If $P_1$ is to send something to $P_2$ at
color class $c$, then $P_2$ will know this at the end of the
pre-communication and it will wait an incoming message from $P_1$ at color
class $c$. As expected, there will be an overhead for this process. We
investigate the gained time by piggybacking as well as the overhead
time incurred by pre-communication in the experiments section.

\begin{figure}
  \centering
  \scalebox{.5}{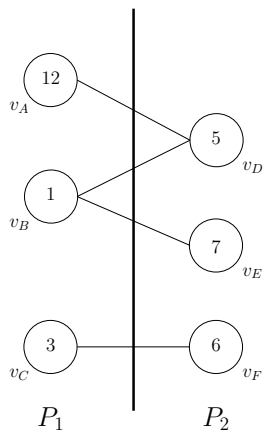}
  \caption{Piggybacking example. Initial colors are written in the vertex.}
  \label{fig:piggy}
\end{figure}

Figure~\ref{fig:piggy} presents the situation between two processors
with a total of six boundary vertices after initial coloring. Vertices that are
not on the boundary between processors $P_1$ and $P_2$ are not
depicted. Without the use of piggybacking, $P_1$ sends three messages
to $P_2$ containing the information of a single vertex at the end of
steps 1, 3 and 12. It also needs to send empty messages during the
other steps. $P_2$ needs to send similar three messages. Using
piggybacking, $P_1$ sends a message to $P_2$ at the end of step 4; the
message contains both the color information of vertices $v_B$ and
$v_C$. Then $P_2$ sends the color information of vertex
$v_D$. Finally, $P_1$ will send the color information of vertex $v_A$
at the end of the recoloring iteration and $P_2$ will send the color
information of both $v_E$ and $v_F$. Notice that in this setting no
empty messages are exchanged. Piggybacking is able to remove all empty
messages and furthermore reduces the total number of non-empty
messages from 6 to 4 between $P_1$ and $P_2$.

\subsection{Random X-Fit}

We expect that synchronous recoloring will be scalable in terms of
number of colors. Piggybacking will make recoloring scalable in terms
of runtime, especially if the number of vertices per color is
roughly balanced.

So, we investigate a fast coloring alternative for the initial
coloring. In~\cite{BGMBC-jpdc}, Bozda\u{g} et al. proposed two
different color selection strategies; First Fit and Staggered First
Fit. We propose another color selection strategy, Random-X Fit, which
uniformly selects a random color from the first X available
colors~\cite{Gebremedhin02paralleldistance-k}. We expect that Random-X
Fit will significantly reduce the number of conflicts in the coloring
procedure leading to a fast coloring. Moreover, it should balance the
number of vertices in each color which should help the recoloring
procedure being fast.

Random-X Fit is expected to give solutions with high number of colors. But, we
believe that recoloring is strong enough to fix any bad number of
colors obtained in the initial coloring. And the coordinated use of
Random-X Fit and recoloring should make the procedure fast.

\section{Experiments}
\label{sec:expe}

In the experiments we study the
effectiveness of the recoloring procedure in
Section~\ref{subsec:recoloring} and prove three points:
\begin{itemize}
\item recoloring can significantly improved solution of low quality in
  a few iterations (Section~\ref{subsubsec:seq})
\item our piggybacking techniques makes recoloring scale much better
  (Section~\ref{sec:improved_comm_exp})
\item and in a distributed memory setting, recoloring improves
  significantly the quality of the solution and incurs a low runtime
  cost allowing to use multiple recoloring iterations
  (Section~\ref{sec:dist}).
\end{itemize}
Then we show in Section~\ref{sec:usingrand} that using a Random X-Fit
coloring leads to a low quality solution but, when coupled with
recoloring, it leads to better quality solution in a smaller amount of
time than using a first fit coloring. This allows us to study the
time-quality trade-off of the coloring problem and we exhibit two sets
of parameters that can be used to obtain either a fast solution
``speed'' or a good solution ``quality''.

Before showing experimental results, we present how the experiments are
conducted in Section~\ref{sec:expesetting}.

\subsection{Experimental Setting}
\label{sec:expesetting}

We implemented all the algorithms in Zoltan~\cite{zoltanug_v3}, an
MPI-based C library for parallel partitioning, load balancing,
coloring and data management services for distributed-memory systems.
For the graphs, we partitioned them by ParMETIS~\cite{parmetis31} version 3.1.1 or with
a simple block partitioning onto the parallel platform.

All of our experiments are performed on an in-house cluster with
64 computing nodes. There are two Intel Xeon
E5520 (quad-core clocked at 2.27GHz) processors at each node along with
48GB of main memory, and 500
GB of local hard disk. Interconnection between the nodes are done through 20Gbps DDR
oversubscribed InfiniBand. Nodes run CentOS with the Linux kernel
2.6.32. The C code is compiled with GCC 4.1.2 using the -O2
optimization flag. We tried different MPI implementations and
MVAPICH2 in version 1.6 is chosen to utilize
the InfiniBand interconnect efficiently.

Experiments are performed on number of processors which are powers of
2 from 1 to 512 processors. Each physical machine has 8
cores. When we do the allocation of processors on the cluster, we first
use processors on different nodes to highlight distributed memory
issue. That means, when 64 processors are used, each allocated processor (core)
is on a different machine. Using 128 processors allocates 2 cores per
machine and an allocation of 512 processors uses the 8 cores of the 64
machines. We have shown in~\cite{HiPC11} that this setting is the fastest configuration.

We run the experiments on six real-world application graphs and three
synthetically generated graphs. The real-world graphs are from different
application areas including linear car analysis, finite element,
structural engineering and automotive industry~\cite{GM00,SH04}. We
obtained them from the University of Florida Sparse Matrix
Collection\footnote{\url{http://www.cise.ufl.edu/research/sparse/matrices/}}
and the Parasol project. Table~\ref{tab:prop_real} gives the list of
the graphs and their main properties. We also listed the number of
colors obtained with a sequential run of the three vertex-visit
orderings in the table. Lastly, runtime of
the {\em Natural} coloring in sequential is given in table. One important thing
to note is that the biggest real-world graph takes less than half a second to color sequentially
using a {\em Natural} ordering, which means that the distributed coloring of all the
graphs very challenging.

\begin{table}
  \centering
  \begin{tabular}{|l|r|r|r|r|r|r|r|} \hline
    \multicolumn{1}{|c|}{Name} & \multicolumn{1}{|c|}{$\card{V}$} & \multicolumn{1}{|c|}{$\card{E}$} & \multicolumn{1}{|c|}{$\maxdegree$} & \multicolumn{1}{|c|}{NAT} & \multicolumn{1}{|c|}{LF} & \multicolumn{1}{|c|}{SL} &\multicolumn{1}{|c|}{seq. time}\\\hline
    auto    & 448,695    &  3,314,611 &  37 & 13&12&10&0.1103s\\
    bmw3\_2 & 227,362    &  5,530,634 & 335 & 48&48&37&0.0836s\\
    hood    & 220,542    &  4,837,440 &  76 & 40&39&34&0.0752s\\
    ldoor   & 952,203    & 20,770,807 &  76 & 42&42&34&0.3307s\\
    msdoor  & 415,863    &  9,378,650 &  76 & 42&42&35&0.1458s\\
    pwtk    & 217,918    &  5,653,257 & 179 & 48&42&33&0.0820s\\ \hline
  \end{tabular}
  \caption{Properties of real-world graphs.}
  \label{tab:prop_real}
\end{table}

\begin{table}
  \centering
  \begin{tabular}{|l|r|r|r|r|r|r|} \hline
    \multicolumn{1}{|c|}{Name} & \multicolumn{1}{|c|}{$\card{V}$} & \multicolumn{1}{|c|}{$\card{E}$} & \multicolumn{1}{|c|}{$\maxdegree$} & \multicolumn{1}{|c|}{NAT} & \multicolumn{1}{|c|}{LF} & \multicolumn{1}{|c|}{SL}\\\hline
    ER   & 16,777,216 & 134,217,624 &     42 & 12  & 10 & 10\\
    Good & 16,777,216 & 134,181,065 &  1,278 & 28  & 15 & 14\\
    Bad  & 16,777,216 & 133,658,199 & 38,143 & 146 & 89 & 88 \\ \hline
  \end{tabular}
  \caption{Properties of synthetic graphs.}
  \label{tab:prop_rmat}
\end{table}

The synthetically generated graphs are RMAT graphs,
introduced by Chakrabarti et al.~\cite{RMAT}. 
When generating the RMAT graphs, the adjacency matrix of 
the graph is subdivided in 4 equal parts and edges are
distributed to these parts with a given probabilities.
Specifying different probabilities
allows the generation of different classes of random graphs. 
We generated three graphs, RMAT-ER, RMAT-Good and RMAT-Bad. Their degree
distributions for the four parts are $(0.25, 0.25, 0.25, 0.25)$, 
$(0.45, 0.15, 0.15, 0.25)$ and $(0.55, 0.15, 0.15, 0.15)$, respectively.
We generate these three graphs to create variety of challenges
for our distributed memory algorithms. For instance, RMAT-ER is
in the class of Erd\H os-R\'enyi random graphs which are known to be hard to
partition. Other two graphs are scale-free graphs with small-world
property and power-law degree distribution. This class of graphs
are also difficult to partition and easily create very unbalanced workloads
if a vertex-partitioning scheme is used. Properties of those graphs
are given in Table~\ref{tab:prop_rmat}. In distributed experiments, 
we partition these graphs using a block partitioning.

For all the experiments we will present both the number of colors
obtained and the runtime of the method when the number of processors
varies. The real-world graphs all show the same trends and their
results are aggregated in the following manner. Each value (number of
colors and runtime) is first normalized with respect to the value
obtained by the {\em Natural} ordering of the same graph on one
processor. Then the normalized value for different graphs are
aggregated using a geometric mean. The three randomly generated graphs
are presented independently for number of colors and they will be aggregated
for runtime results, which are normalized with respect to {\em Natural}
ordering on 4 processors, since they show the same behavior in terms of runtime.

\subsection{Recoloring}
\label{subsec:recoloring}

\subsubsection{Sequential Setting}
\label{subsubsec:seq}

\begin{figure}[bt]
  \centering
  \subfigure[Natural (NAT)]{\includegraphics[width=.32\linewidth,page=1]{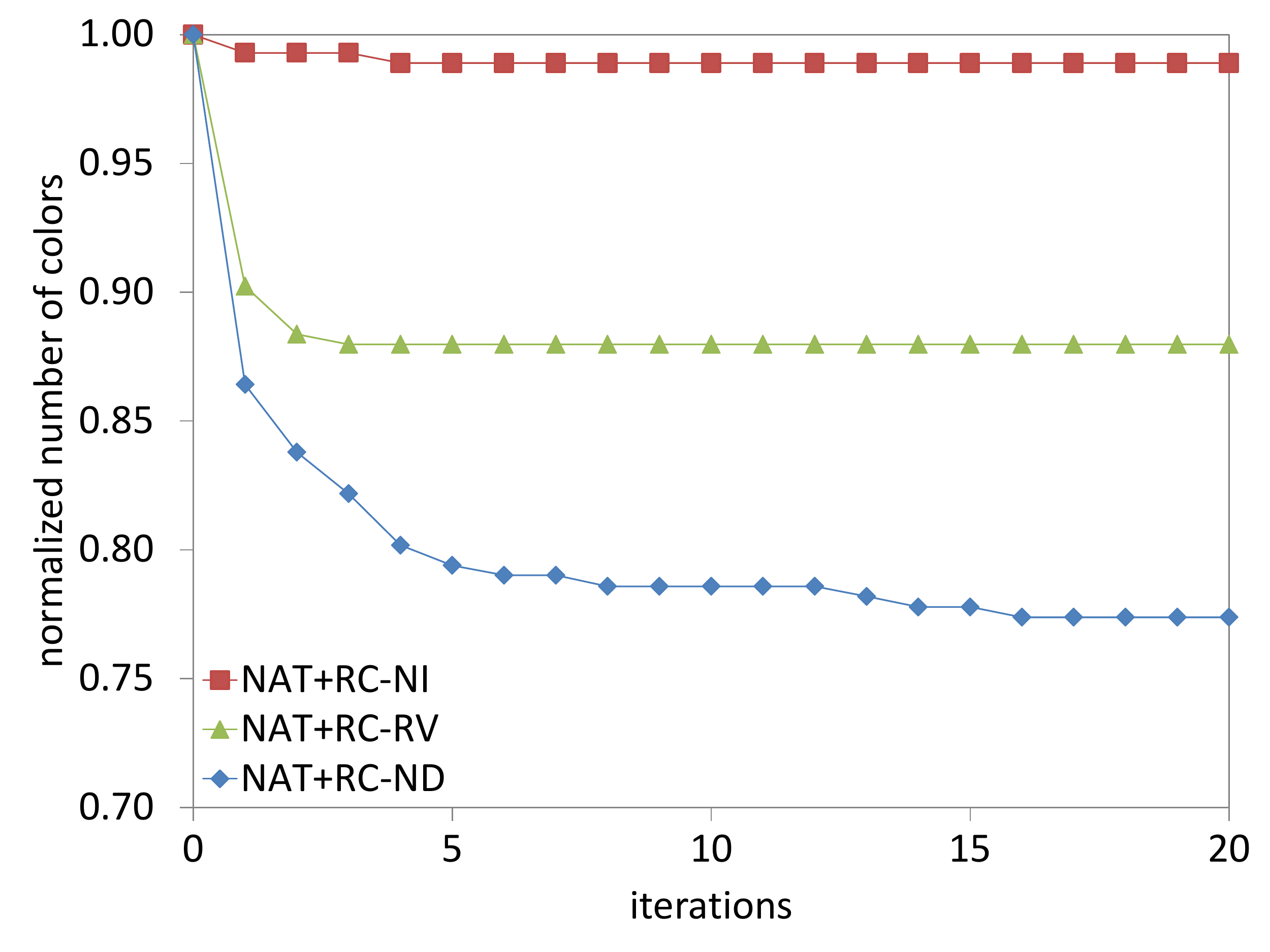}\label{fig:seq_study_nat}}
  \subfigure[Largest First (LF)]{\includegraphics[width=.32\linewidth,page=2]{charts/iter.pdf}\label{fig:seq_study_lf}}
  \subfigure[Smallest Last (SL)]{\includegraphics[width=.32\linewidth,page=3]{charts/iter.pdf}\label{fig:seq_study_sl}}
  \caption{Sequential study of recoloring with different vertex-visit orderings on the real-world graphs~\cite{HiPC11}.}
  \label{fig:seq_study}
\end{figure}

In Figure~\ref{fig:seq_study}, we present the effect of vertex-visit orderings and
multiple iterations of recoloring on the number of colors in sequential settings, which
was also presented in~\cite{HiPC11}. Different vertex-visit orderings are multiplexed
with different color permutation strategies in each chart. For example. NAT+RC-ND gives
the {\em Natural} ordering coupled with recoloring with the Non-Decreasing number of
vertex order of the color classes. Note that the charts begin with
0 iteration which shows the quality of the vertex-visit ordering only.

The first important result in those charts is that the LF vertex ordering results in
lower number of colors than the {\em Natural} ordering with $0.96$ normalized number of
colors without recoloring and SL is the best vertex ordering strategy with $0.78$ normalized
number of colors. The second result is that three tested permutations of the color
classes result in a decrease of the number of colors. Among them, the NI permutation
leads to smallest improvement whereas the ND permutation gives the smallest number of colors
by obtaining $0.8$ normalized number of colors after $20$ iterations for all three orderings.
Lastly, the best number of colors is obtained by combining the SL vertex ordering with
ND color permutation (SL+RC-ND). ND permutation leads to lower number of colors due to
the selection of color classes. Success of a permutation can be measured by its ability to
remove as many color classes as possible. In ND permutation strategy, color classes with
fewer vertices are selected first so that the classes with larger number of vertices 
can merge with them. 

In this paper, we investigate the effect of randomness of the
color class permutations in the sequential setting to get out of the
local plateaus. For this purpose, we
run experiments on real-world graphs with the three vertex-visit orderings,
namely Natural, Largest First and Smallest Last. We choose the ND
permutation as a reference point since it gives the best number of
colors. The color permutation that picks a random permutation uniformly is denoted
RAND. Using the RAND permutation at every $x$ iterations and using ND
for the other iterations is denoted (ND-RAND\%$x$). For instance using RAND
every five iterations is denoted (ND-RAND\%5) while using it every 10
iterations is denoted (ND-RAND\%10). We also investigate the variant
we call (ND-RAND\%$2^i$) which uses random permutation only at iterations
which are power of two; in other words the ND permutation is used at
each iteration except at iterations 2, 4, 8, 16, ... where the
permutation is RAND.

\begin{figure}[bt]
  \centering
  \subfigure[Natural (NAT)]{\includegraphics[width=.32\linewidth,page=1]{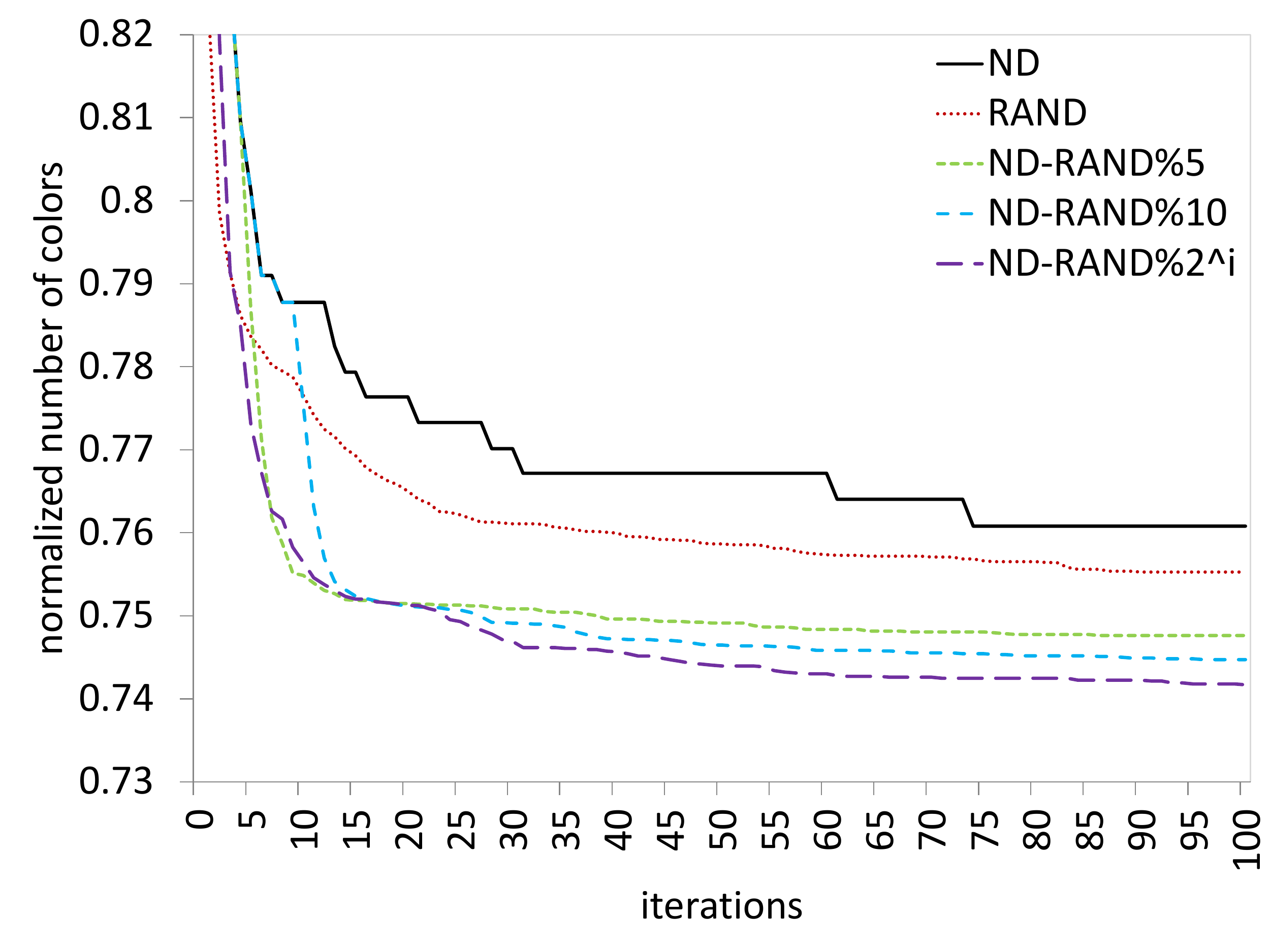}\label{fig:random_perm_nat}}
  \subfigure[Largest First (LF)]{\includegraphics[width=.32\linewidth,page=2]{charts/random_perm.pdf}\label{fig:random_perm_lf}}
  \subfigure[Smallest Last (SL)]{\includegraphics[width=.32\linewidth,page=3]{charts/random_perm.pdf}\label{fig:random_perm_sl}}
  \caption{Sequential study of recoloring with color class permutation randomness on the real-world graphs.}
  \label{fig:random_perm}
\end{figure}

Figure~\ref{fig:random_perm} presents the results for this
experiment. Results are obtained by running all the tests 10
times and taking the average over all runs. For the NAT vertex-visit ordering, randomness
helps reducing the number of colors. The RAND permutation is better
than the ND permutation for all number of iterations. However,
rarefied randomness outperforms pure randomness. As the frequency
of randomness decreases, the number of colors decreases as
well. For example, ND-RAND\%10 is better than ND-RAND\%5. Two main
conclusions can be drawn from this experiment. First, randomization
helps getting ND out of local plateau as is strongly suggested by the
sudden decrease in number of colors at the 10th iteration of
ND-RAND\%10. Second, ND might need numerous iterations to refine a
coloring: the normalized number of colors does not change between
iteration 32 and 60 of ND. The
ND-RAND\%$2^i$ manages to successfully use these two properties by
trying to get out of local plateau aggressively during the first
iterations and then, by letting ND to slowly refine its coloring.

On the other hand, the LF and SL orderings show different behaviors
than the NAT ordering. Randomness helps reducing the number of colors during
the first iterations (up to 20 iterations in our experiments). But the
ND permutation outperforms all of the other permutations on large number of
iterations.  This picture suggests that insisting on the ND permutation
for the LF and SL vertex-visit orderings gives the best results if one is
interested on high number of iterations. Overall, it can be said that randomness
provides a sharp decrease in number of colors at the beginning of the process, but then the number of colors converges 
rapidly. Since we start from a lower number of colors in LF and SL orderings,
it is expected to converge quicker and this is the case in LF and SL orderings
with high number iterations of random permutations.

Figure~\ref{fig:seq_study} showed that the SL+RC-ND combination
outperformed all other vertex-visit orderings and color class
permutations presented in the sequential case. We will then focus on
comparing the distributed synchronous recoloring (RC) and the
asynchronous one (aRC) using the non recolored {\em Smallest Last} ordering
as a reference. However, before diving into more comparison, we
need to investigate the effect of the piggybacking technique on the
communication scheme of synchronous recoloring procedure.

\subsubsection{Improvements on Communication Scheme of Recoloring}
\label{sec:improved_comm_exp}

\begin{figure}
  \centering
  \includegraphics[width=.491\linewidth]{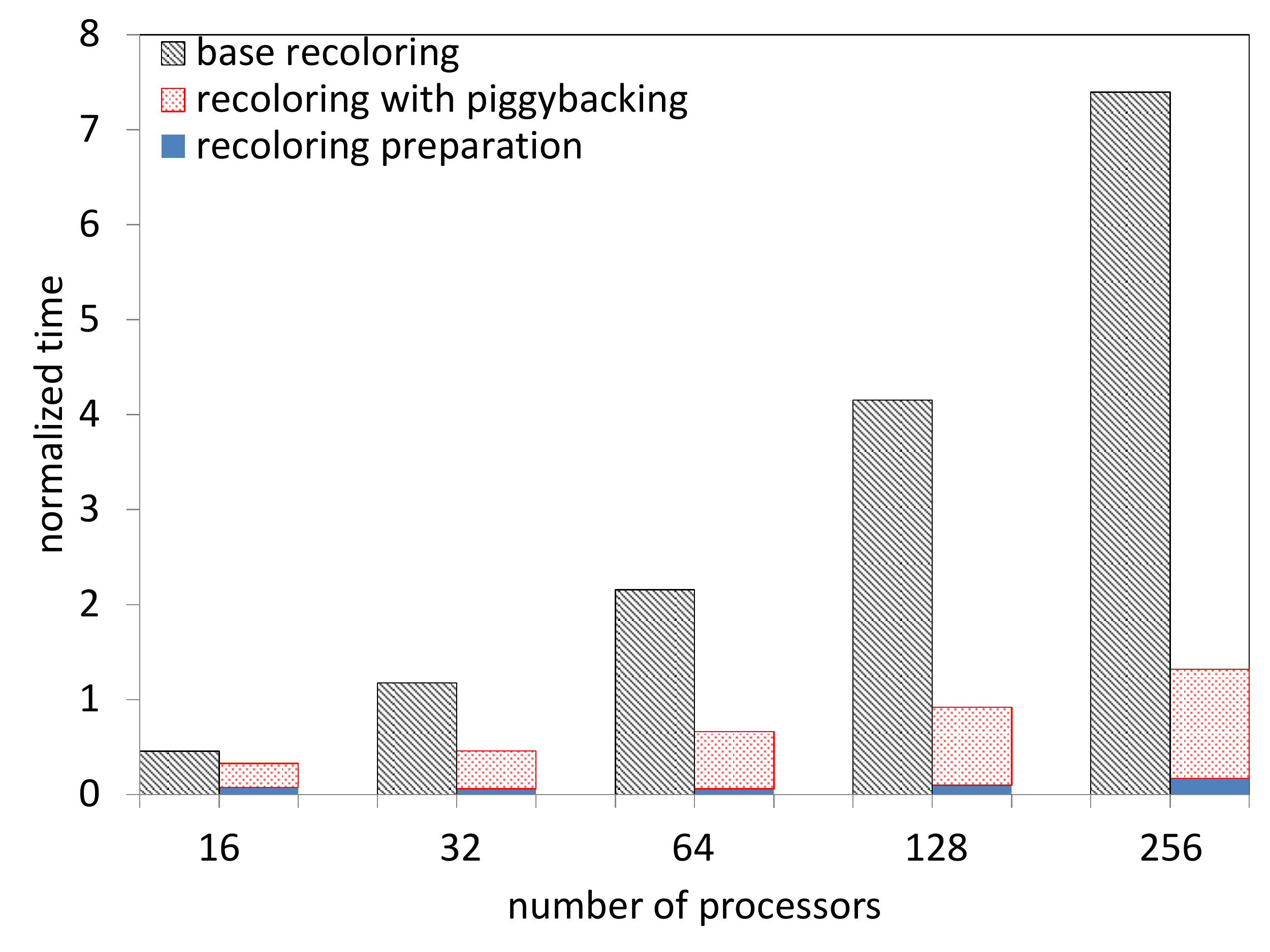}
  \caption{Comparison of one recoloring iteration of base recoloring in~\cite{HiPC11} and improved scheme on the real-world graphs.}
  \label{fig:reduced_comm}
\end{figure}

We show here the runtime improvement obtained on recoloring by applying the piggybacking\
techniques discussed in Section~\ref{sec:improved_comm}. We run the experiments with 8 processors per node 
configuration to see the impact easily. Figure~\ref{fig:reduced_comm} presents
a comparison between a base implementation of recoloring (without piggybacking presented in~\cite{HiPC11})
and the improved implementation using piggybacking
with detailed timings. Using piggybacking needs a preparation phase, but it takes at most 12\%
of the improved total coloring time in the worst case. On the other hand, piggybacking provides
a huge improvement in recoloring time. Our experiments show that piggybacking provides
80\% less number of messages in average, when compared to base implementation. Improved coloring with piggybacking gives 20\% to 70\% improvement over base total
coloring time. These results indicate that accumulating messages into bigger chunks provides a
better communication for recoloring and we will then use the 
improved recoloring technique when we are making
runtime comparisons.

\begin{figure}
  \centering
  \subfigure{\includegraphics[width=.491\linewidth,page=1]{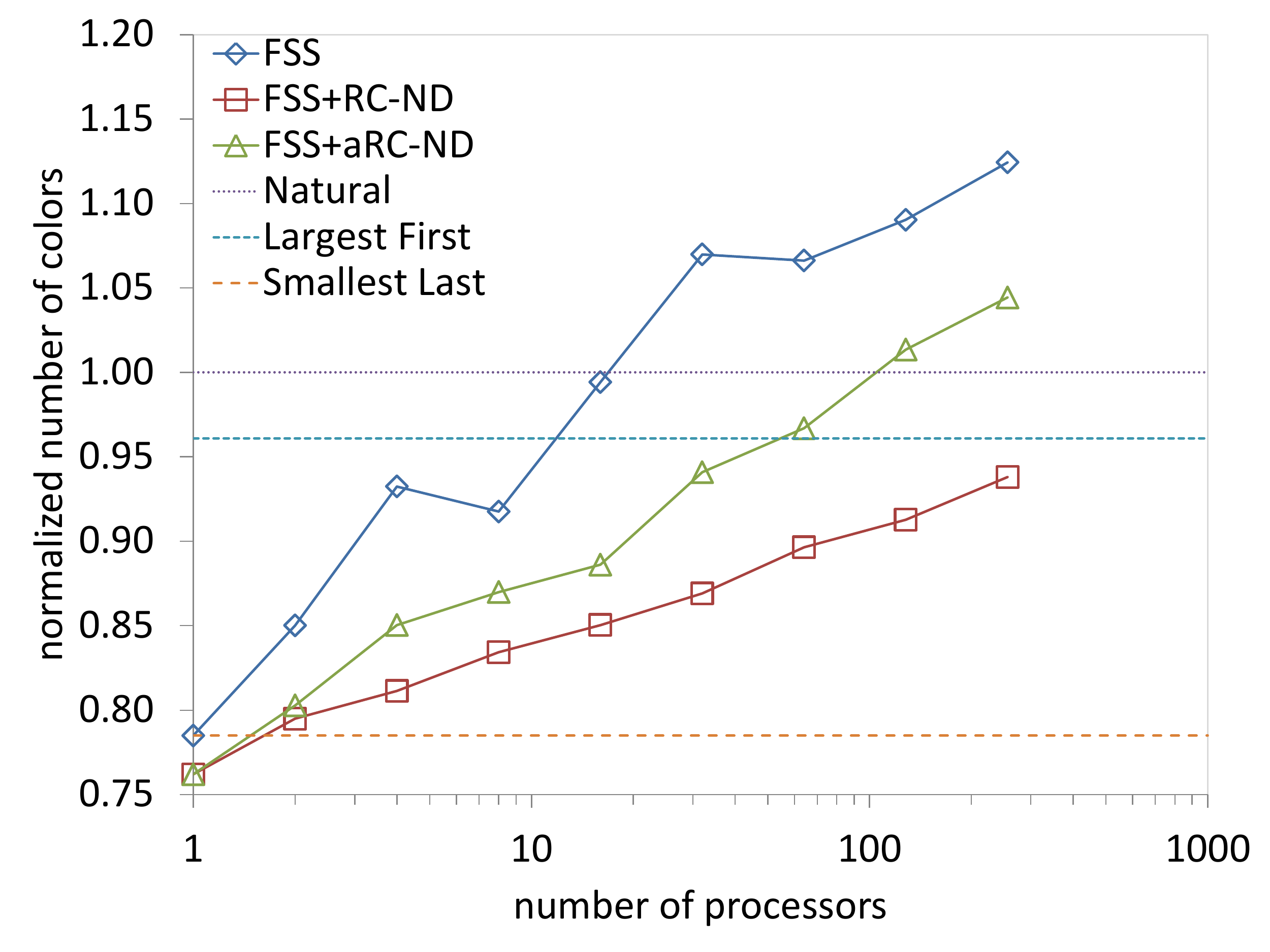}}
  \subfigure{\includegraphics[width=.491\linewidth,page=2]{charts/real_worlds_rc.pdf}}
  \caption{Comparison of recoloring on real-world graphs with Smallest Last ordering.}
  \label{fig:comp_recolor_real}
\end{figure}

\subsubsection{Distributed Setting}
\label{sec:dist}

Figure~\ref{fig:comp_recolor_real} presents the performance of
recoloring on the real-world graphs. We compare the combination of First-Fit color selection,
{\em Smallest Last} ordering and synchronous communication (denoted FSS as in, ~\cite{HiPC11}), with
additional synchronous and asynchronous recoloring.
As expected, the synchronous
recoloring reduces the normalized number of colors significantly
providing to keep the normalized number of colors below the one
obtained using the sequential {\em Largest First} even on 256 processors,
bringing a 18\% improvement in the number of colors to FSS. However the synchronous recoloring takes more
time than the FSS coloring, reaching a normalized runtime of
$2.01$ while FSS has a normalized runtime of $0.60$ on 256 processors. 
Asynchronous recoloring provides a middle ground by
allowing to obtain a better quality coloring than FSS is able to
achieve. However, asynchronous recoloring gives almost the same runtime
with synchronous recoloring. It shows a different picture than the one 
in~\cite{HiPC11} since the piggybacking technique in synchronous recoloring brings significant improvement.
Synchronous recoloring is as fast as the asynchronous recoloring,
and leads to higher quality solutions.

\begin{figure}[tbh!]
  \centering
  \subfigure[Number of colors on RMAT-ER]{\includegraphics[width=.491\linewidth,page=3]{charts/RMATs_rc.pdf}\label{fig:comp_rc_ER}}
  \subfigure[Number of colors on RMAT-Good]{\includegraphics[width=.491\linewidth,page=2]{charts/RMATs_rc.pdf}\label{fig:comp_rc_Good}}
  \subfigure[Number of colors on RMAT-Bad]{\includegraphics[width=.491\linewidth,page=1]{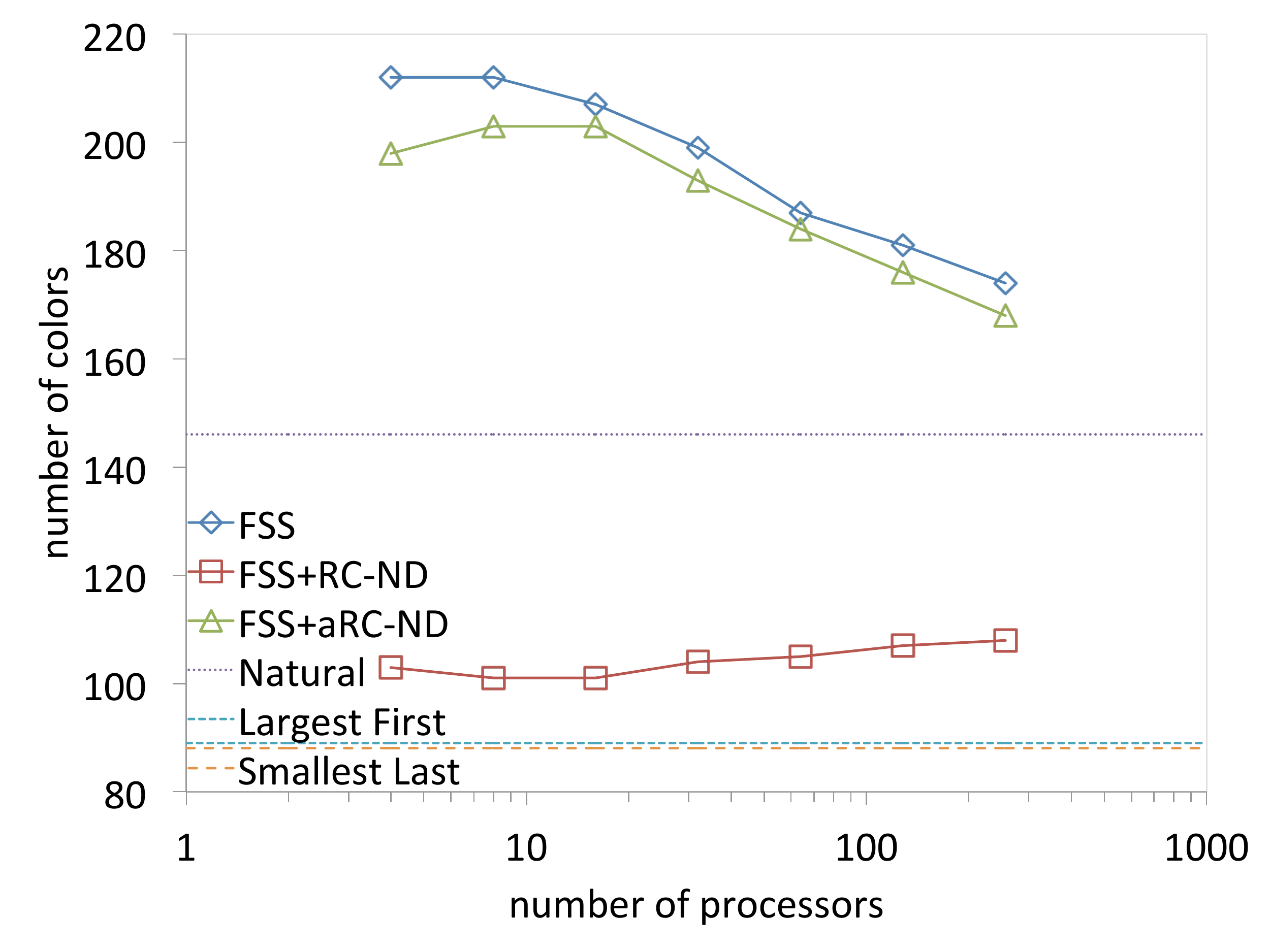}\label{fig:comp_rc_Bad}}
  \subfigure[Runtime]{\includegraphics[width=.491\linewidth,page=4]{charts/RMATs_rc.pdf}\label{fig:comp_rc_time}}
  \caption{Impact of recoloring on RMAT graphs.}
\end{figure}

The impact of recoloring for number of colors is presented on the randomly generated graphs
in Figures~\ref{fig:comp_rc_ER},~\ref{fig:comp_rc_Good}, and ~\ref{fig:comp_rc_Bad}.
Figure~\ref{fig:comp_rc_time} shows the aggregated runtime results for RMAT graphs.
The number of colors of FSS was mainly given by the number of conflicts it
generated. The vertex-visit ordering computed by the asynchronous
recoloring does not avoid the majority of these conflicts and the
improvement in number of colors compared to FSS is less than 10\%.

Synchronous recoloring shows a different picture. While FSS used
many colors because of a high number of conflicts on RMAT-Good
and RMAT-Bad, the synchronous recoloring does not yield any
conflict. Therefore, it obtains a much better number of colors, close
to the sequential {\em Largest First} and {\em Smallest Last} orderings
with at most 50\% improvement compared to FSS. In terms of runtime, the absence of
conflict and the size of the graphs make the synchronous recoloring
procedure very scalable, inducing a very low overhead
compared to the initial coloring when the number of processors is
high.

\begin{figure}
  \centering
  \includegraphics[width=.60\linewidth]{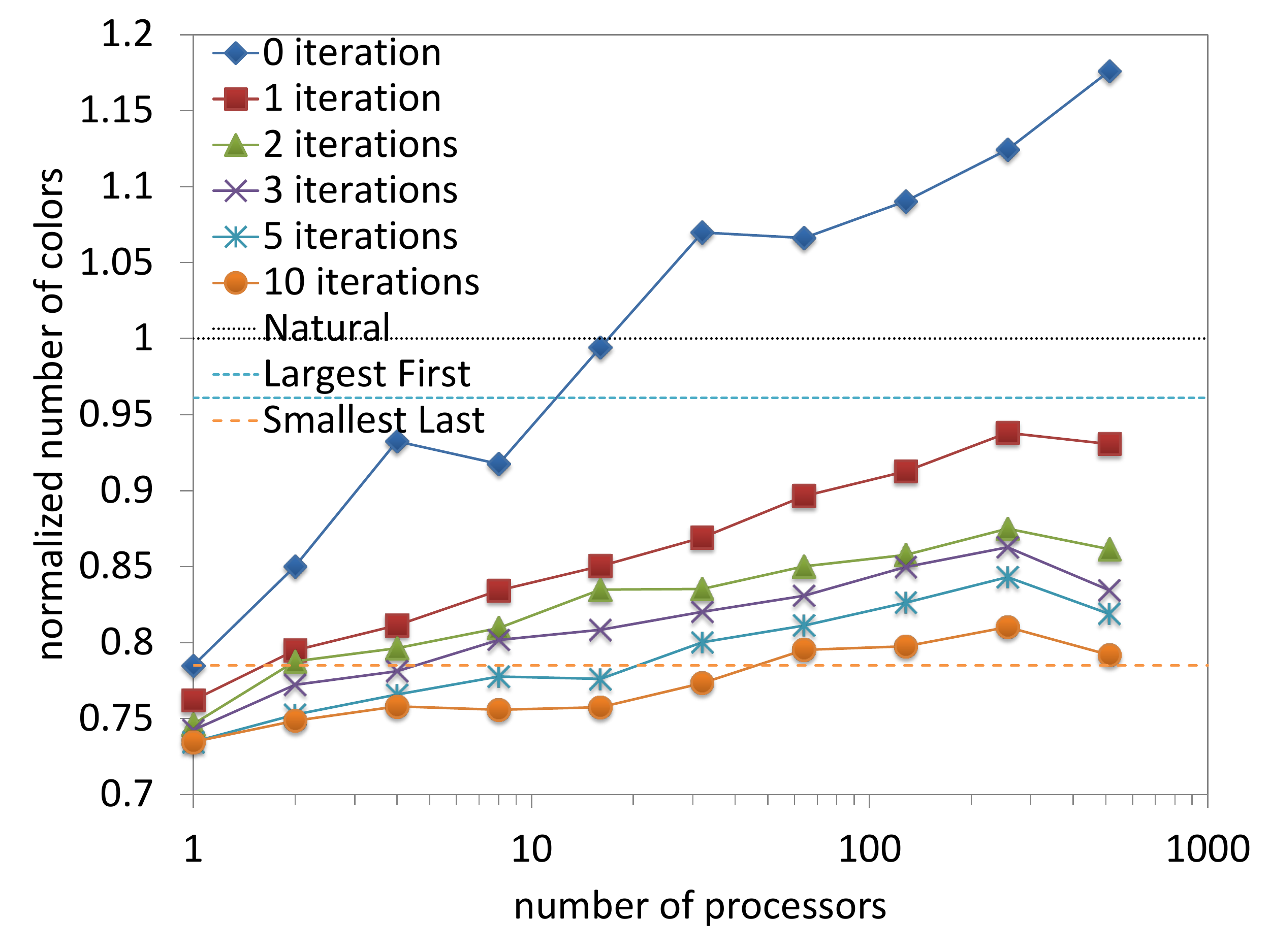}
  \caption{Impact of the number of recoloring iterations on real-world graphs
     in distributed memory~\cite{HiPC11}.}
  \label{fig:comp_recolor_iter_dist}
\end{figure}

Given the benefit of the recoloring procedure, it is of the interest
to catch the sequential improvement on the number of colors obtained
by performing the recoloring procedure multiple times. The effect of multiple
recoloring iterations in distributed-memory settings on the real-world
graphs is given in Figure~\ref{fig:comp_recolor_iter_dist}.
Single iteration of recoloring provides significant gain over no recoloring
by staying below sequential {\em Largest First} when number of processors
is 512. On the other hand, running 10 iterations of the recoloring
procedure on 512 processors provides very good number of colors, which is close
to the sequential {\em Smallest Last} vertex-visit ordering.

\subsection{Using Random-X Fit with Recoloring}
\label{sec:usingrand}

\begin{figure}[tb]
  \centering
  \subfigure{\includegraphics[width=.6\textwidth,page=1]{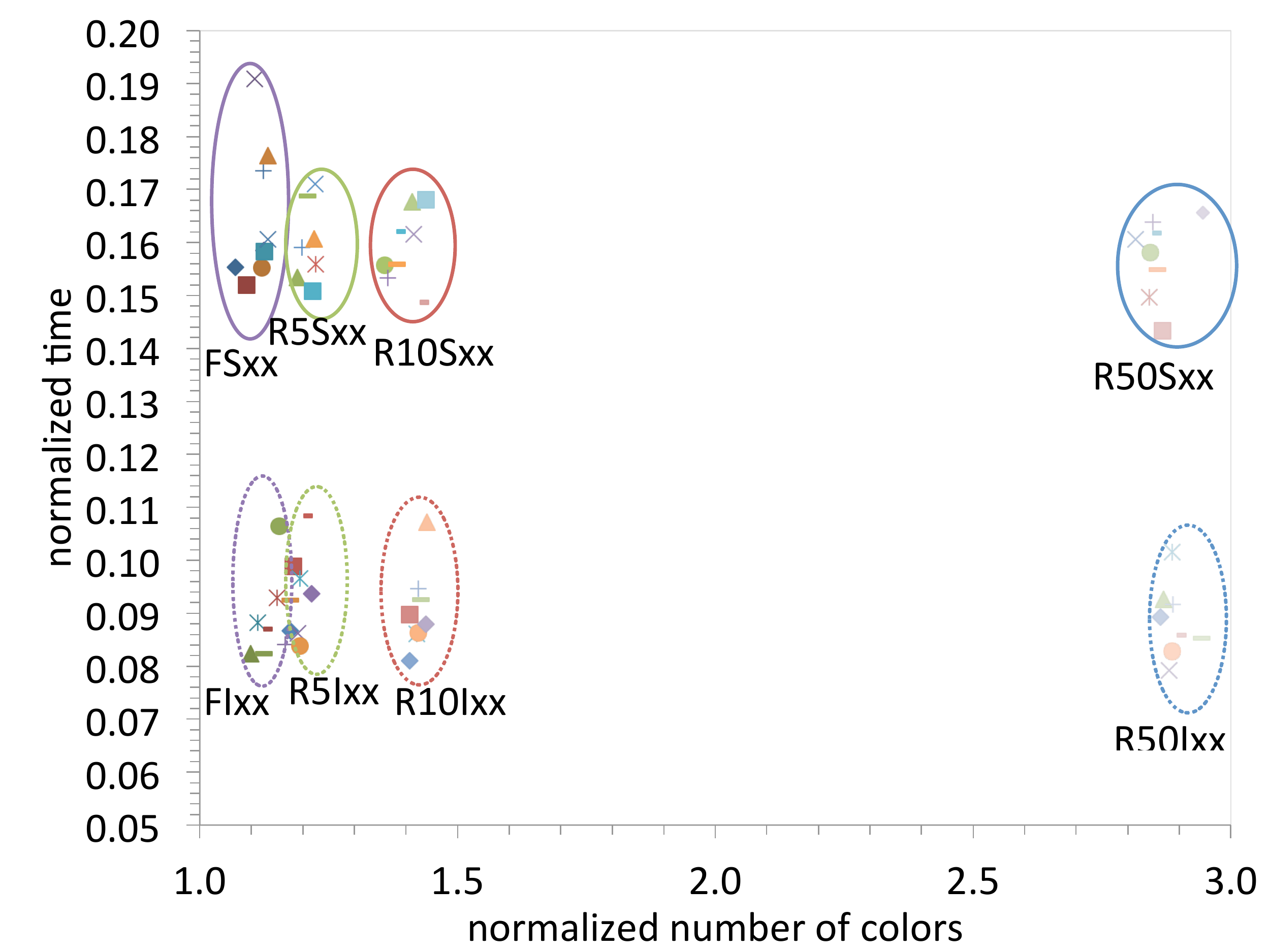}}
  \caption{Comparison of different combinations of parameters on real-world graphs in terms of number of colors and runtime
without recoloring.}
  \label{fig:random1}
\end{figure}

We investigate the effect of Random-5 Fit, Random-10 Fit and Random-50
Fit within our coloring framework with different number of recoloring
iterations. We believe that Random-X Fit can provide an initial
solution that is fast and that balances the number of vertex per color
which is very suitable for recoloring. Various number of processors
were tried, but they lead to similar results, so we only present
results using 32 processes on 32 nodes.

Firstly, we investigate the original coloring with Random-X Fit color
selection strategy. We run experiments with varying and combining all
of the parameters we have: superstep size (500, 1000, 5000 and 10000),
vertex-visit ordering (Internal First and Smallest Last), synchronous
and asynchronous communication patterns for original coloring and
color selection strategies (First Fit, Random-5 Fit, Random-10 Fit and
Random-50 Fit). Quality and runtime results are presented in
Figure~\ref{fig:random1}.  Note that, superstep size and communication
pattern does not have a significant impact on quality-runtime
trade-off, so we clustered the nearer points and tagged them with
respect to their color selection strategy and vertex-visit
ordering. For example, R5Ixx stands for Random-5 Fit with Internal
First vertex-visit ordering where superstep size and communication
pattern vary.  As expected, Internal First ordering gives better
runtime results than SL ordering while the situation is the other way
around for the number of colors. As the X factor in Random-X Fit
strategy increases, the number of colors degrades since the selection
is done from a larger set.

\begin{figure}[tbh!]
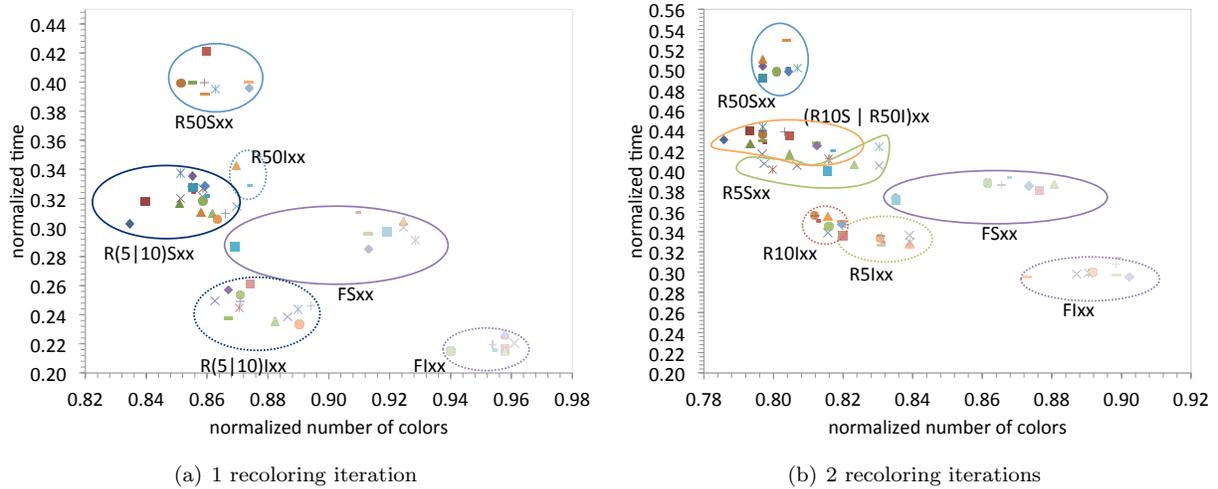

  \centering
  \subfigure[1 recoloring iteration]{\includegraphics[width=.491\linewidth,page=2]{charts/random_coloring.pdf}\label{fig:random2}}
  \subfigure[2 recoloring iterations]{\includegraphics[width=.491\linewidth,page=3]{charts/random_coloring.pdf}\label{fig:random3}}
  \caption{Comparison of different combinations of parameters on real-world graphs in terms of number of colors and runtime
with Non-Decreasing permutation recoloring.}
\end{figure}

Next, we investigate the use of Random-X color selection strategy
in the context of recoloring. Results for one recoloring iteration
are shown in Figure~\ref{fig:random2}.  All
Random-X color selection strategies give better number of colors than
First Fit with one recoloring iteration. It means that although
Random-X color selection strategies start recoloring with higher
number of colors than First Fit, they result in less colors at the end
of the recoloring. In our opinion, this result is very important and
shows that random color selection strategies provide perfect ground
for recoloring.  However, Random-X color selection strategies are not
as good as First Fit in terms of runtime. Since the runtime of
recoloring is correlated with the number of colors at the beginning of
recoloring procedure, the recoloring procedure after Random-X Fit
strategy takes longer. Despite this fact, R(5|10)Ixx combinations give
better number of colors and better runtime than FSxx combinations,
which were suggested as the best combination for recoloring
in~\cite{HiPC11}. We also investigate the impact of Random-X color selection
strategies with two recoloring iterations. Figure~\ref{fig:random3}
presents the results for this experiment.  Results are similar to the
ones presented in Figure~\ref{fig:random2}.

\begin{figure}[tb]
  \centering
  \subfigure{\includegraphics[width=.6\linewidth,page=4]{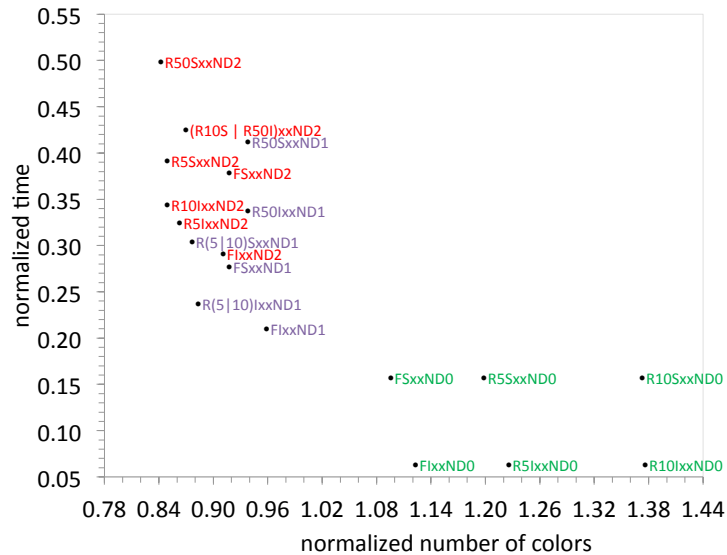}}
  \caption{Comparison of different combination of parameters up to two Non-Decreasing permutation recoloring iterations.}
  \label{fig:bigpicture}
\end{figure}

The trade-off between number of colors and runtime is presented in
Figure~\ref{fig:bigpicture} where results with 0,1 and 2 recoloring
iterations are combined. Notice that, in Figure~\ref{fig:bigpicture},
R(5|10)IxxND1 is better than FIxxND2 and FSxxND2 both in terms of
number of colors and runtime.  In summary, the random color selection
strategy compensates the impact of vertex-visit ordering and the best
combination to be suggested becomes R(5|10)IxxND1, which is using the
Random-5 or Random-10 color selection strategy with Internal First
Ordering and one Non-Decreasing recoloring iteration. This allows us
to identify two interesting sets of parameters. This makes
R(5|10)IxxND1 a good candidate for having a solution of good quality
and we call this set of parameters ``quality''.

For a obtaining a solution quickly, it is better not to use
recoloring. The random strategies tend to significantly hinders the
quality of the solution in this case and using First Fit is leading to
a higher quality solution at no runtime overhead. The local ordering
of the vertices makes little difference in quality because of the
distributed execution of algorithm; therefore the fastest ordering
(Internal first) is preferred. This makes FIxxND0 a good candidate for
obtaining quickly a solution of moderate quality and we call this set
of parameters ``speed''.

\section{Conclusion}
\label{sec:ccl}

In this paper we investigated different ways of
improving the number of colors in distributed-memory graph coloring
algorithms. We investigated recoloring by
utilizing the independent sets of vertices resulted by an existing
solution to color large number of vertices independently with little
synchronization or with little conflicts. We showed that graphs that
result in a larger number of conflicts benefit significantly from synchronous
recoloring in terms of the number of colors. The runtime overhead
is being small in such cases, therefore multiple iterations of recoloring can be
used to obtain even fewer colors. We also showed
that synchronous recoloring can be implemented using piggybacking to
improve network access patterns; making synchronous recoloring as fast
as the previously proposed asynchronous recoloring.
Overall, recoloring proved to be a scalable concept
for the number of colors when number of processors increases, thanks
to the equivalence to a sequential procedure.
Last but may be most interestingly, we showed that the Random-X Fit color 
selection strategy works quite well with recoloring and provides
better number of colors and runtime results than the vertex-visit
ordering solutions.

Overall, we suggest
two combinations of the parameters; if the user is interested in a quick
coloring with decent quality, then the ``speed'' policy FIxxND0 (First Fit, Internal first, no recoloring)
is suitable; if the user needs better coloring, then the ``quality'' R(5|10)IxxND1
combination of Random-X Fit, with Internal First ordering and some
Non-Decreasing recoloring iterations is best for large number of processors. 

As a future work, more thorough study of Random-X Fit strategy
could lead to a better understanding of the coloring problem and could
suggest ways to derandomize this color selection strategy.

\section*{Acknowledgments}
This work was partially supported by the U.S. Department of Energy SciDAC Grant
DE-FC02-06ER2775 and NSF grants CNS-0643969, OCI-0904809 and OCI-0904802.

\bibliographystyle{abbrv}
\bibliography{paper}

\begin{thebibliography}{10}

\bibitem{ABC94}
J.~Allwright, R.~Bordawekar, P.~D. Coddington, K.~Dincer, and C.~Martin.
\newblock A comparison of parallel graph coloring algorithms.
\newblock Technical Report SCCS-666, Northeast Parallel Architectures Center at
  Syracuse University (NPAC), 1994.

\bibitem{zoltanug_v3}
E.~Boman, K.~Devine, R.~Heaphy, B.~Hendrickson, V.~Leung, L.~A. Riesen,
  C.~Vaughan, {\"U}.~{\c{C}}ataly\"urek, D.~Bozda\u{g}, W.~Mitchell, and
  J.~Teresco.
\newblock {\em {Zoltan 3.0}: Parallel Partitioning, Load Balancing, and
  Data-Management Services; User's Guide}.
\newblock Sandia National Laboratories, Albuquerque, NM, 2007.
\newblock Tech. Report SAND2007-4748W.

\bibitem{BGMBC-jpdc}
D.~Bozda\u{g}, A.~Gebremedhin, F.~Manne, E.~Boman, and
  {\"{U}}.~{\c{C}}ataly{\"u}rek.
\newblock A framework for scalable greedy coloring on distributed memory
  parallel computers.
\newblock {\em Journal of Parallel and Distributed Computing}, 68(4):515--535,
  2008.

\bibitem{Brelaz1979}
D.~Br\'{e}laz.
\newblock New methods to color the vertices of a graph.
\newblock {\em Commun. ACM}, 22:251--256, April 1979.

\bibitem{Chaitin}
G.~J. Chaitin.
\newblock Register allocation \& spilling via graph coloring.
\newblock {\em SIGPLAN Not.}, 17:98--101, June 1982.

\bibitem{CM83}
T.~F. Coleman and J.~J. More.
\newblock Estimation of sparse {J}acobian matrices and graph coloring problems.
\newblock {\em SIAM Journal on Numerical Analysis}, 1(20):187--209, 1983.

\bibitem{Culberson92iteratedgreedy}
J.~C. Culberson.
\newblock Iterated greedy graph coloring and the difficulty landscape.
\newblock Technical Report TR 92-07, University of Alberta, June 1992.

\bibitem{RMAT}
Y.~Z. D.~Chakrabarti and C.~Faloutsos.
\newblock R-{M}{A}{T}: A recursive model for graph mining.
\newblock In {\em Proceedings of SIAM Data Mining}, 2004.

\bibitem{dubr81}
R.~D. Dutton and R.~C. Brigham.
\newblock A new graph colouring algorithm.
\newblock {\em The Computer Journal}, 24(1):85--86, 1981.

\bibitem{Ellis:1989:LVG:74142.74153}
J.~A. Ellis and P.~M. Lepolesa.
\newblock A las vegas graph colouring algorithm.
\newblock {\em The Computer Journal}, 32:474--476, Oct. 1989.

\bibitem{gamst_freq}
A.~Gamst.
\newblock Some lower bounds for a class of frequency assignment problems.
\newblock {\em IEEE Transactions on Vehicular Technology}, 35(1):8--14, Feb.
  1986.

\bibitem{Garey_circuit}
M.~Garey, D.~Johnson, and H.~So.
\newblock An application of graph coloring to printed circuit testing.
\newblock {\em IEEE Transactions on Circuits and Systems}, 23(10):591--599,
  Oct. 1976.

\bibitem{Gebremedhin_parallelgraph}
A.~H. Gebremedhin and F.~Manne.
\newblock Parallel graph coloring algorithms using {OpenMP} (extended
  abstract).
\newblock In {\em In First European Workshop on {OpenMP}}, pages 10--18, 1999.

\bibitem{GM00}
A.~H. Gebremedhin and F.~Manne.
\newblock Scalable parallel graph coloring algorithms.
\newblock {\em Concurrency: Practice and Experience}, 12:1131--1146, 2000.

\bibitem{Gebremedhin02paralleldistance-k}
A.~H. Gebremedhin, F.~Manne, and A.~Pothen.
\newblock Parallel distance-k coloring algorithms for numerical optimization.
\newblock In {\em Euro-Par 2002 Parallel Processing - 8th International
  Conference}, pages 912--921, 2002.

\bibitem{GMP05}
A.~H. Gebremedhin, F.~Manne, and A.~Pothen.
\newblock What color is your jacobian? {G}raph coloring for computing
  derivatives.
\newblock {\em SIAM Review}, 47(4):629--705, 2005.

\bibitem{JGT:JGT3190080115}
M.~K. Goldberg.
\newblock Edge-coloring of multigraphs: Recoloring technique.
\newblock {\em Journal of Graph Theory}, 8(1):123--137, 1984.

\bibitem{Jones94}
M.~T. Jones and P.~E. Plassmann.
\newblock Scalable iterative solution of sparse linear systems.
\newblock {\em Parallel Computing}, 20:753--773, 1994.

\bibitem{parmetis31}
G.~Karypis, K.~Schloegel, and V.~Kumar.
\newblock {ParMETIS}: Parallel graph partitioning and sparse matrix ordering
  library, version 3.1.
\newblock Technical report, Dept. Computer Science, University of Minnesota,
  2003.

\bibitem{kosow}
A.~V. Kosowski and K.~Manuszewski.
\newblock Classical coloring of graphs.
\newblock {\em Graph Colorings}, pages 1 -- 19, 2004.

\bibitem{Manne98aparallel}
F.~Manne.
\newblock A parallel algorithm for computing the extremal eigenvalues of very
  large sparse matrices.
\newblock In {\em Lecture Notes in Computer Science}, pages 332--336, 1998.

\bibitem{matula_SL}
D.~W. Matula.
\newblock A min-max theorem for graphs with application to graph coloring.
\newblock {\em SIAM Review}, 10:481--482, 1968.

\bibitem{Matula1983}
D.~W. Matula and L.~L. Beck.
\newblock Smallest-last ordering and clustering and graph coloring algorithms.
\newblock {\em J. ACM}, 30:417--427, July 1983.

\bibitem{matula_72}
D.~W. Matula, G.~Marble, and J.~Isaacson.
\newblock Graph coloring algorithms.
\newblock {\em Graph Theory and Computing}, pages 109--122, 1972.

\bibitem{Saad99}
Y.~Saad.
\newblock Ilum: A multi-elimination ilu preconditioner for general sparse
  matrices.
\newblock {\em SIAM Journal of Scientific Computing}, 17:830--847, 1999.

\bibitem{HiPC11}
A.~E. Sar{\i}y\"uce, E.~Saule, and {\"{U}}.~V. {\c{C}}ataly\"urek.
\newblock Improving graph coloring on distributed-memory parallel computers.
\newblock In {\em 18th International Conference on High Performance Computing
  (HiPC)}, 2011.

\bibitem{Strout02}
M.~M. Strout, L.~Carter, J.~Ferrante, J.~Freeman, and B.~Kreaseck.
\newblock Combining performance aspects of irregular gauss-seidel via sparse
  tiling.
\newblock In {\em 15th Workshop on Languages and Compilers for Parallel
  Computing (LCPC)}, 2002.

\bibitem{SH04}
M.~M. Strout and P.~D. Hovland.
\newblock Metrics and models for reordering transformations.
\newblock In {\em Proceedings of the The Second {ACM SIGPLAN} Workshop on
  Memory System Performance (MSP)}, pages 23--34, June 8 2004.

\bibitem{Turner:1988:AKC:48880.48884}
J.~S. Turner.
\newblock Almost all $k$-colorable graphs are easy to color.
\newblock {\em J. Algorithms}, 9:63--82, Mar. 1988.

\bibitem{Welsh01011967}
D.~J.~A. Welsh and M.~B. Powell.
\newblock An upper bound for the chromatic number of a graph and its
  application to timetabling problems.
\newblock {\em The Computer Journal}, 10(1):85--86, 1967.

\bibitem{Zuckerman}
D.~Zuckerman.
\newblock Linear degree extractors and the inapproximability of max clique and
  chromatic number.
\newblock {\em Theory of Computing}, 3:103--128, 2007.

\end{thebibliography}

\end{document}